# NEW OBSERVATIONS AND A NEW INTERPRETATION OF CO(3–2) IN IRAS F 10214+4724

D. Downes

Institut de Radio Astronomie Millimétrique, 38406 St. Martin d'Hères, France

P. M. Solomon

Astronomy Program, State University of New York, Stony Brook, NY 11794

and

S. J. E. Radford

National Radio Astronomy Observatory, Tucson AZ 85721-0665

## ABSTRACT

New observations with the IRAM interferometer of CO(3–2) from the highly luminous galaxy IRAS F 10214+4724 show the source is $1.5'' \times \leq 0.9''$ ; they display no evidence of any velocity gradient. This size, together with optical and IR data that show the galaxy is probably gravitationally lensed, lead to a new model for the CO distribution. In contrast to many lensed objects, we have a good estimate of the intrinsic CO and far IR surface brightnesses, so we can derive the CO and far IR/sub-mm magnifications. The CO is magnified 10 times and has a true radius of 400 pc and the far IR is magnified 13 times and has a radius of 250 pc. The true far IR luminosity is $4 - 7 \times 10^{12}\,L_\odot$ and the molecular gas mass is $2 \times 10^{10}\,M_\odot$. This is nearly an order of magnitude less than previously estimated. Because the far IR magnification is lower than the mid and near IR magnification, the intrinsic spectral energy distribution now peaks in the far infrared. That is, nearly all of the energy of this object is absorbed and re-emitted in the far infrared. In CO luminosity, molecular gas content, CO linewidth, and corrected far IR luminosity, 10214+472 is a typical, warm, IR ultraluminous galaxy.

## 1. INTRODUCTION

This paper reports new observations with the IRAM interferometer of CO(3–2) from the highly luminous galaxy IRAS F 10214+4724 at $z = 2.3$. Earlier observations with this instrument showed a source size of $2.5'' \times 1.0''$ with an east-west velocity gradient (Radford, Brown, & Vanden Bout 1993). The interferometer has since been improved by the installation of a fourth antenna, better SIS mixer receivers, and a new broad-band spectral correlator, so we repeated the observations to check the earlier results. The new data indicate the source is somewhat smaller than deduced previously — $1.5''\times \leq 0.9''$ — and show no sign of any velocity shift across the source. This result, together with optical data that show the source is probably gravitationally lensed (Broadhurst & Lehar 1995; Graham & Liu 1995; Trentham 1995; Eisenhardt et al. 1995; Serjeant et al. 1995), lead to a re-interpretation of the CO distribution, a determination of the CO magnification, and a lower estimate of the molecular mass and far IR luminosity, all of which show that 10214+4724 is a warm ultraluminous IR galaxy.

## 2. OBSERVATIONS AND RESULTS

The observations were made with the four 15 m antennas of the IRAM interferometer on Plateau de Bure, France, in 1993 and 1994. Three different configurations gave 18 baselines from 24 to 288 m, and a beam of $2.3'' \times 1.6''$ (FWHM). Typical receiver temperatures were 80 K at 105 GHz. The spectral correlator covered 1300 km s$^{-1}$ at 7 km s$^{-1}$ resolution, but for analysis we smoothed the data to resolutions of 20, 40, and 80 km s$^{-1}$. Amplitude and phase were calibrated with the 5.5 Jy source 0923+392, whose flux was measured relative to strong quasars and planets with both the interferometer and the IRAM 30 m telescope. Table 1 summarizes our results. CO(3–2) from 10214+4724 was detected over a 400 km s$^{-1}$ interval. In the integrated line intensity map (Fig. 1), the source contours run east-west, while the beam is oriented at 42° E of N. This suggests the source is resolved. To avoid deconvolving the synthesized beam, we fit a Gaussian model directly to the interferometer visibilities. This fit indicates the source size is $1.5'' \times \leq 0.9''$ (FWHM) at $84° \pm 12°$ E of N. A similar fit to the phase calibrator, 0923+392, yielded a unresolved point source $< 0.3''$ in size. In $80\,\mathrm{km\,s^{-1}}$ wide channel maps (Fig. 1), the source also has the same size in the line center and in the line wings. We find no evidence for any velocity gradient. Neither fits to individual channel visibilities nor position – velocity diagrams show any position shift greater than $0.3''$ relative to the central position anywhere in the central $160\,\mathrm{km\,s^{-1}}$ of the CO(3–2) line. The CO(3–2) flux of 10214+4724 is $4.1 \pm 0.8\,\mathrm{Jy\,km\,s^{-1}}$. This corresponds to an apparent CO luminosity[1] of $4.6 \times 10^{10}\,\mathrm{K\,km\,s^{-1}\,pc^2}$.

In the interferometer CO(3–2) spectrum (Fig. 2), the peak intensity of $17 \pm 1.5$ mJy and the linewidth of $220\,\mathrm{km\,s^{-1}}$ are nearly the same as the 18.5 mJy and $230\,\mathrm{km\,s^{-1}}$ we measured with the 30 m telescope (Solomon et al. 1992). Table 1 summarizes the observations. Within the uncertainties, the redshift $z = 2.2854 \pm 0.0001$ agrees with previously determined CO redshifts. It differs slightly from the $2.284 \pm 0.001$ determined for H$\alpha$, [N II], and [O III] (Elston et al. 1994), possibly indicating the CO occupies a different volume than the optical Narrow Line Region. Outside the line channels, there is no continuum source at the position of 10214+4724 or anywhere else in the $45''$ field of view. At the exact position of 10214+4724, the limit is 1 mJy. The non-detection is consistent with our previous measurement of 10 mJy at 1.2 mm and an optically thin dust spectrum (Downes et al. 1992), with no contribution from a non-thermal continuum.

---

[1]We use $H_0 = 75\,\mathrm{km\,s^{-1}\,Mpc^{-1}}$ and $q_0 = 0.5$ throughout this paper.

## 3. DISCUSSION

In our initial interpretation of CO in 10214+4724, we argued the CO(3–2), (4–3), and (6–5) line ratios indicate molecular clouds at a kinetic temperature of 50 K and an $H_2$ density of 5000 $cm^{-3}$. We estimated a total mass of $2 \times 10^{11}$ $M_\odot$ from the apparent CO(3–2) luminosity (Solomon et al. 1992). Now we propose a new model (Table 2) for the CO distribution, motivated by the small extent of the CO emission and the optical and 2.2 $\mu$m evidence for gravitational lensing (Matthews et al. 1994; Graham & Liu 1995; Broadhurst & Lehar 1995; Serjeant et al. 1995; Eisenhardt et al. 1995). Since the lens magnification depends on the source's true size and precise location, different components may have different magnifications. As a result, not only the luminosity but also the colors of 10214+4724 may be changed by the lens. This differential magnification can be estimated from observed sizes and models of the thermal CO and far IR emission.

### 3.1. Magnification of the CO Image

The 2.2 $\mu$m images (Matthews et al. 1994; Graham & Liu 1995) show a compact 0.7″ diameter source superposed on a weaker arc 1.5″ long. This morphology strongly suggests a gravitational lens with at least two components. The IRAM interferometer has poorer resolution than the 2.2 $\mu$m image so it is not possible to directly trace CO in the thin arc. Hence we convolved the 2.2 $\mu$m image with the interferometer beam to see if it would reproduce the CO image. These simulations show the CO cannot follow the 2.2 $\mu$m distribution of a compact source superposed on a weaker arc; this produces a much more compact CO image than observed. On the other hand, a convolution of the interferometer beam with the arc alone, truncated above the lower 10% to 20% contour level of the 2.2 $\mu$m image, fits the CO(3–2) image very well. Subtracting this model from the data leaves a residual near zero with an r.m.s. noise level similar to the data. This suggests the observed $1.5''$ CO source coincides with the $2.2\,\mu$m arc. The model image is compared with the CO data in Plate 1, which also shows the $2.2\,\mu$m image and the arc used in the convolution.

Since the intrinsic CO brightness temperature can be estimated from line ratios, we can determine an upper limit to the lens magnification. The gravitational lens magnifies the source in one dimension, preserving surface brightness. The smallest possible intrinsic CO source is an optically thick sphere or disk radiating thermally over the linewidth $\Delta V$. The observed CO luminosity

$$L'_{\rm CO}({\rm obs}) \quad \equiv \quad m_{\rm CO}\, \pi\, R^2\, T_b\, f_V\, \Delta V \quad , \tag{1}$$

where $R$ is the true source radius, $T_b$ is its rest frame brightness temperature, $f_V$ is the velocity filling factor, and $m_{\rm CO}$ is the source magnification. We approximate the arc segment by an ellipse, so $m_{\rm CO} = a/R$, where $a = D_A\, \theta_{\rm obs}/2$ is the apparent semi-major axis of the magnified CO source, $\theta_{\rm obs}$ is the observed angular length of the arc segment, and $D_A$ is the angular size distance. At $z = 2.28$, $D_A/\theta = 5.3\,{\rm kpc\ arcsec^{-1}}$. Then

$$m_{\rm CO} \quad = \quad \frac{a}{R} \quad = \quad \frac{\pi\, a^2\, \Delta V}{L'_{\rm CO}({\rm obs})}\, f_V\, T_b \quad . \tag{2}$$

A magnification factor that varies with the *square* of the apparent source size and *inversely* with the apparent luminosity may seem counter-intuitive at first glance. Since apparent luminosity is proportional to $a\,R$, however, the magnification is actually only linearly proportional to apparent size. In equation (2), lens magnification is expressed in terms of measurable quantities, $\Delta V$, $L'_{\rm CO}$, and $T_b$.

$T_b$ can be estimated from the observed CO(6–5)/CO(3–2) ratio (Solomon et al. 1992) which clearly indicates warm molecular gas. For a gas kinetic temperature of 60 K, slightly less than the dust temperature of 80 K, we find that an LVG model that fits the observed CO line intensities and ratios has CO(3–2) and (6–5) brightness temperatures of $43 \pm 7$ and $27 \pm 5$ K, respectively. The opacities are 6 in CO(1–0), 37 in CO(3–2), and 41 in CO(6–5).

For the observed CO luminosity and linewidth (Table 1), and for $T_b = 43\,\mathrm{K}$,

$$m_{\rm CO} = 4.6\ \theta^2_{\rm obs}\, f_V\, (T_b/43\,\mathrm{K}) \quad . \tag{3}$$

with $\theta_{\rm obs}$ in arcsec. The observed CO major axis is $1.5''$, so for $f_V \leq 1.0$,

$$m_{\rm CO}\ \leq\ 10\,(T_b/43\,\mathrm{K}) \quad . \tag{4}$$

Even after allowing for this modest lens magnification, the CO luminosity is still high, comparable to the most CO rich luminous IR galaxies as suggested by Graham & Liu (1995). The intrinsic CO(3–2) luminosity of 10214+4724 is $L'_{\rm CO}({\rm true})\ \geq\ 5 \times 10^9\,\mathrm{K\,km\,s^{-1}\,pc^2}$ with a true CO radius $R\ \geq\ 400\,\mathrm{pc}$. For comparison, in a sample of 37 ultraluminous IR galaxies out to $z = 0.27$, the highest CO luminosity is 16 $\times 10^9$ $\mathrm{K\,km\,s^{-1}\,pc^2}$ and the average is $7 \times 10^9$ $\mathrm{K\,km\,s^{-1}\,pc^2}$ (Solomon et al. 1996). The molecular mass implied by these observations and the lens model is 5 – 10 times lower than previously estimated, removing the apparent conflict between the molecular and dynamical masses. The total $H_2$ mass is $2 \times 10^{10}\,\mathrm{M_\odot}$ for $M(\mathrm{H_2})/L'_{\rm CO} = 4\,\mathrm{M_\odot(K\,km\,s^{-1}\,pc^2)^{-1}}$ (Radford, Solomon, & Downes 1991). For the CO true radius of $400/f_V$ pc, the observed linewidth of $220\,\mathrm{km\,s^{-1}}$ , and an inclination of 45°, the dynamical mass $M_{\rm dyn} = 3 \times 10^{10}\, f_V\ \mathrm{M_\odot}$. Hence molecular gas makes up a large fraction of the total mass of 10214+4724, as it does in low redshift ultraluminous IR galaxies.

### 3.2. Far-IR and Mid-IR Magnification

Most of the observed bolometric luminosity of the source comes from two components that contribute equally to the observed spectral energy $\nu\, S_\nu$ . The first, observed at 450 to 1300 $\mu$m, is far IR emission by optically thick dust at $\sim$ 80 K (Downes et al. 1992). The second component, the IRAS 60 $\mu$m flux, is rest frame 18 $\mu$m mid IR emission from hotter dust at about 200 K. Without any downward correction for lens magnification, each component has an apparent luminosity of $7\times10^{13}\,L_\odot$.

In Broadhurst & Lehar's interpretation of the Keck 2.2 $\mu$m and WHT images, the galaxy's emission is magnified 50 times to give the observed luminosity. This magnification would mean the true radius of the 0.7″ core is 40 pc, a typical radius for an AGN Narrow Line Region (NLR). While this may be also appropriate for the mid-IR emission, it is much too small to account for the thermal far IR emission at 80 K.

As with the CO, we can estimate the far IR magnification because we know the dust temperature, $T_d$, from the shape of the far IR/sub-mm continuum spectrum. For a black body, the apparent luminosity

$$\frac{L_{\rm fir}({\rm obs})}{L_\odot} = m_{\rm fir}\left(\frac{R_{\rm bb}}{R_\odot}\right)^2\left(\frac{T_d}{T_\odot}\right)^4 \quad , \tag{5}$$

where $m_{\rm fir}$ is the far IR magnification and $R_{\rm bb}$ is the true black body radius. If $a_{\rm fir}$ is the apparent semi-major axis of the magnified far IR source, then

$$m_{\rm fir} \equiv \frac{a_{\rm fir}}{R_{\rm bb}} = 1.78\,\frac{a_{\rm fir}^2\,T_d^4}{L_{\rm fir}({\rm obs})} \tag{6}$$

with $a_{\rm fir}$ in pc, $T_d$ in K, and $L_{\rm fir}$ in $L_\odot$. Substituting $T_d = 80$ K and the observed far IR luminosity of $7\times10^{13}\,L_\odot$,

$$m_{\rm fir} = 7.3\ \ \theta_{\rm fir}^2({\rm obs})\ \left(\frac{T_d}{80\,{\rm K}}\right)^4 \quad , \tag{7}$$

where $\theta_{\rm fir} = 2\,a_{\rm fir}/D_A$ is in arcsec. So far there are no size measurements of the lensed far IR source, but from comparisons with non-lensed ultraluminous galaxies, we find the range is 0.6 – 1.0 times the the true size of the CO source, which we derived to be 400 pc for a CO velocity filling factor of unity. ¿From eqns.(5) and (7), and for $R_{\rm bb} = 250 - 400$ pc and $T_d = 80$ K, the far IR magnification factor $m_{\rm fir} = 10 - 13$. The true far IR luminosity of the source must therefore be $L_{\rm fir}({\rm true}) = (4 - 7)\times10^{12}\,L_\odot$. *Hence the intrinsic far IR emission dominates the mid-IR emission.* The total IR luminosity (mid + far IR) is then $L_{\rm IR}$ = (5 to

8) $\times 10^{12}\,L_{\odot}$. If the CO velocity filling factor is 0.5 – 1.0, then the ratio $L_{\rm IR}/L_{\rm CO} \approx 1000$, similar to warm ultraluminous IR galaxies, albeit at the high end of the distribution.

What about the dust mass? In the unlensed case, we derived an apparent dust mass of $3.5 \times 10^{8}\,M_{\odot}$ from the optically thin continuum flux observed at 1.3 mm. The ratio of apparent gas mass to warm dust masses was 500 (Downes et al. 1992). Since the intrinsic gas and dust masses are, respectively, 10 and 13 times smaller in this new model, their ratio is now slightly higher, 730. A lower limit to the metallicity is $M_{\rm dust}/M_{\rm dyn} = 10^{-3}$. Our original conclusion, based on the similarity of the ratio of the optically thick CO luminosity to the optically thin rest-frame 350 $\mu$m luminosity in 10214+4724 and in several nearby galaxies, remains the same. IRAS 10214+4724 has a heavy element abundance comparable to our Galaxy and nearby galaxies.

### 3.3. Power Source Revisited

The true bolometric luminosity is 10 – 13 times lower and the gas mass is 10 times lower than our earlier estimates (Solomon et al. 1992; Downes et al. 1992). Even with these reductions, i.e., a true luminosity of $4 - 7 \times 10^{12}\,L_{\odot}$ and a gas mass of $2 - 4 \times 10^{10}\,M_{\odot}$ in the central 400 to 800 pc radius, the nature of the energy source remains a problem. IRAS 10214+4724 has a Seyfert 2 spectrum without the usual spectroscopic signatures of star formation (Rowan-Robinson et al. 1991, 1993; Lawrence et al. 1993, Elston et al. 1994; Januzzi et al. 1994; Soifer et al. 1995). Differential magnification, however, may alter the observed optical line spectrum so it no longer represents the intrinsic spectrum. If the galaxy has an extended starburst and an AGN, the small Narrow Line Region will be magnified much more than the larger star-forming ring where the H II regions are. The *minimum* far IR and CO radius is about 400 pc ($0.08''$). This is 10 times larger than the Narrow Line Region, which can be magnified 50 times (Broadhurst & Lehar 1995) rather than 5 – 10 times. In the magnified image, the Narrow Line region appears as the $0.7''$ core. Since the optical spectra are of this core and not the more extended arc, and since the small AGN region would be magnified much more than an extended starburst, the observed spectrum will always be dominated by AGN characteristics.

As with nearby ultraluminous galaxies, the far IR dominates the intrinsic luminosity of 10214+4724, and it must be radiated by an extended region containing a massive molecular interstellar medium, not by a small, tens of parsecs region around an AGN. In CO luminosity, molecular content, and IR luminosity (Table 2), 10214+4724 is a typical ultraluminous IR galaxy. Is the galaxy powered by star formation in the molecular region itself, or are the gas and dust just part of a massive envelope heated by the AGN (Sanders et al. 1989)? The dense molecular gas in these objects makes them prime candidates for huge starbursts. Heckman (1994) notes time scale constraints make a starburst explanation all but impossible for the *apparent* luminosity of 10214+4724. Although more plausible with the smaller intrinsic luminosity and IR luminosity-to-gas-mass ratio $\approx 250\,L_\odot/M_\odot$, a starburst would still require an IMF of high mass stars only and a high star formation efficiency — 20% of all the gas converted to stars in ten million years. Whatever the current power source, the heavy elements, the dust, and the molecular gas must all have been produced by massive stars in the first 1 — 1.5 billion years of the galaxy's existence.

We thank Quanhui Chen for help with the convolutions of the $2.2\,\mu$m image for comparison with the CO map and Jim Barrett for image processing.

This manuscript was prepared with the AAS LaTeX macros v4.0.

Table 1. CO(3–2) Observations of IRAS F 10214+4724 with the IRAM Interferometer.

| | |
|---|---|
| CO Source size (FWHM) | $(1.5'' \pm 0.4'') \times \leq 0.9''$ |
| position angle | $84^\circ \pm 12^\circ$ east of north |
| CO(3–2) position (J2000; $\pm 0.2''$) | $10^{\rm h}24^{\rm m}34.54^{\rm s}$, $47^\circ 09' 09.8''$ |
| CO(3–2) redshift (LSR) | $2.2854 \pm 0.0001$ |
| Luminosity distance, $D_L = (1+z)^2\, D_A$ | 11.7 Gpc ($1'' = 5.3$ kpc) |
| Peak flux density | $17 \pm 1.5$ mJy |
| Line width | $220 \pm 30$ km s$^{-1}$ |
| Integrated line flux, $S\,\Delta V$ | $4.2 \pm 0.8$ Jy km s$^{-1}$ |
| Apparent CO luminosity, $L'_{\rm CO}({\rm obs})$ | $4.6 \times 10^{10}$ K km s$^{-1}$ pc$^2$ |

Table 2. PROPOSED COMPONENTS OF 10214+4724

| | CO | Far IR | mid IR |
|---|---|---|---|
| true radius | $\geq 400\,\mathrm{pc}$ | 250 pc | 40 pc |
| magnification | $\leq 10$ | 13 | 50 |
| magnified diam. | $1.5''$ | $1.0''$ | $0.7''$ |
| | | | |
| temperature | $T_{\rm kin} = 65\,\mathrm{K}$ | $T_{\rm dust} = 80\,\mathrm{K}$ | $T_{\rm dust} = 200\,\mathrm{K}$ |
| true luminosity | $7 \times 10^6\, f_V^{-1}\,\mathrm{L}_\odot$ | $5 \times 10^{12}\,\mathrm{L}_\odot$ | $1 \times 10^{12}\,\mathrm{L}_\odot$ |
| $L'_{\rm CO}$ (true) | $5 \times 10^9\, f_V^{-1}\,\mathrm{L}_\ell$ | — | — |
| | | | |
| *Mass* ($\mathrm{M}_\odot$): | | | |
| dust mass | — | $3 \times 10^7$ | — |
| $H_2$ mass | $2 \times 10^{10}/f_V$ | — | — |
| dynamical mass | $3 \times 10^{10}/f_V$ | — | — |
| | | | |
| remarks | model CO arc | 2.2$\mu$m arc | 2.2 $\mu$m core |
| | (Plate 1-middle ) | (Plate 1-left) | NLR |

Unlensed sizes for $D_L = 11.7\,\mathrm{Gpc}$; $1'' = 5.3\,\mathrm{kpc}$. CO line luminosity unit $\mathrm{L}_\ell \equiv \mathrm{K\,km\,s^{-1}\,pc^2}$

Fig. 1.— *Top and lower left panels:* Maps of CO(3–2) from 10214+4724, in 80 $\mathrm{km\,s^{-1}}$ wide channels centered on the velocity in the upper left of each box, relative to 105.220 GHz. Contour step = 2.6 mJy $\mathrm{beam^{-1}}$, with negative contours dashed, zero contour omitted; r.m.s. noise = 1.4 mJy $\mathrm{beam^{-1}}$.
*Lower right panel:* CO(3–2) integrated intensity in a 400 $\mathrm{km\,s^{-1}}$ wide band centered on 105.22 GHz. Contour step = 1.4 mJy $\mathrm{beam^{-1}}$; r.m.s. noise = 0.8 mJy $\mathrm{beam^{-1}}$.
Coordinates are J2000. The cross marks the CO(3–2) centroid (Table 1). The $2.3'' \times 1.6''$ clean beam is shown in the lower left of the last panel.

Fig. 2.— CO(3–2) spectrum of 10214+4724 obtained with the IRAM interferometer at 20 $\mathrm{km\,s^{-1}}$ resolution. The data were restored with a $3.7'' \times 3''$ beam to include all the source flux. Velocity offsets are relative to 105.220 GHz.

Fig. 3.— **Plate 1:** CO Arc and Image. Coordinates are R.A. and Dec. offsets in arcsec.
*Left panel*: Keck telescope 2.2 $\mu$m image with grey scale and contours at 2, 4, 6, 12, 25, 50, and 75% of the peak. The feature north of the arc is the lensing galaxy and the other two features are unrelated (Matthews et al. 1994).
*Middle panel*: Truncated 2.2 $\mu$m image used to model the CO arc and the half power shape of the IRAM interferometer beam.
*Right panel*: IRAM interferometer CO map in grey scale and dashed contours. Solid contours are the model image obtained by convolving the beam with the arc. The fit is very good and the difference between the observed image and the model has the same r.m.s. noise level as in the original map.

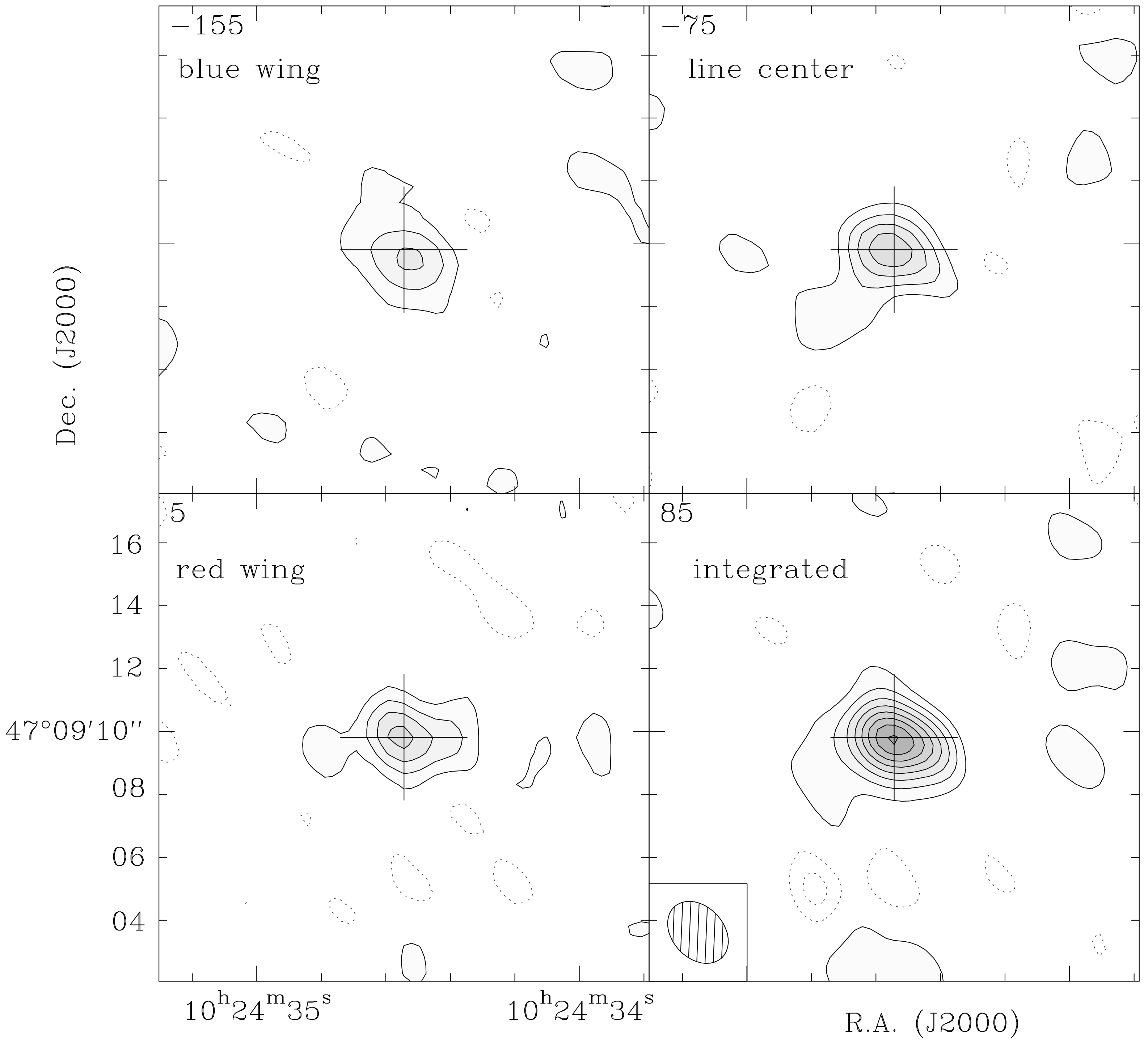

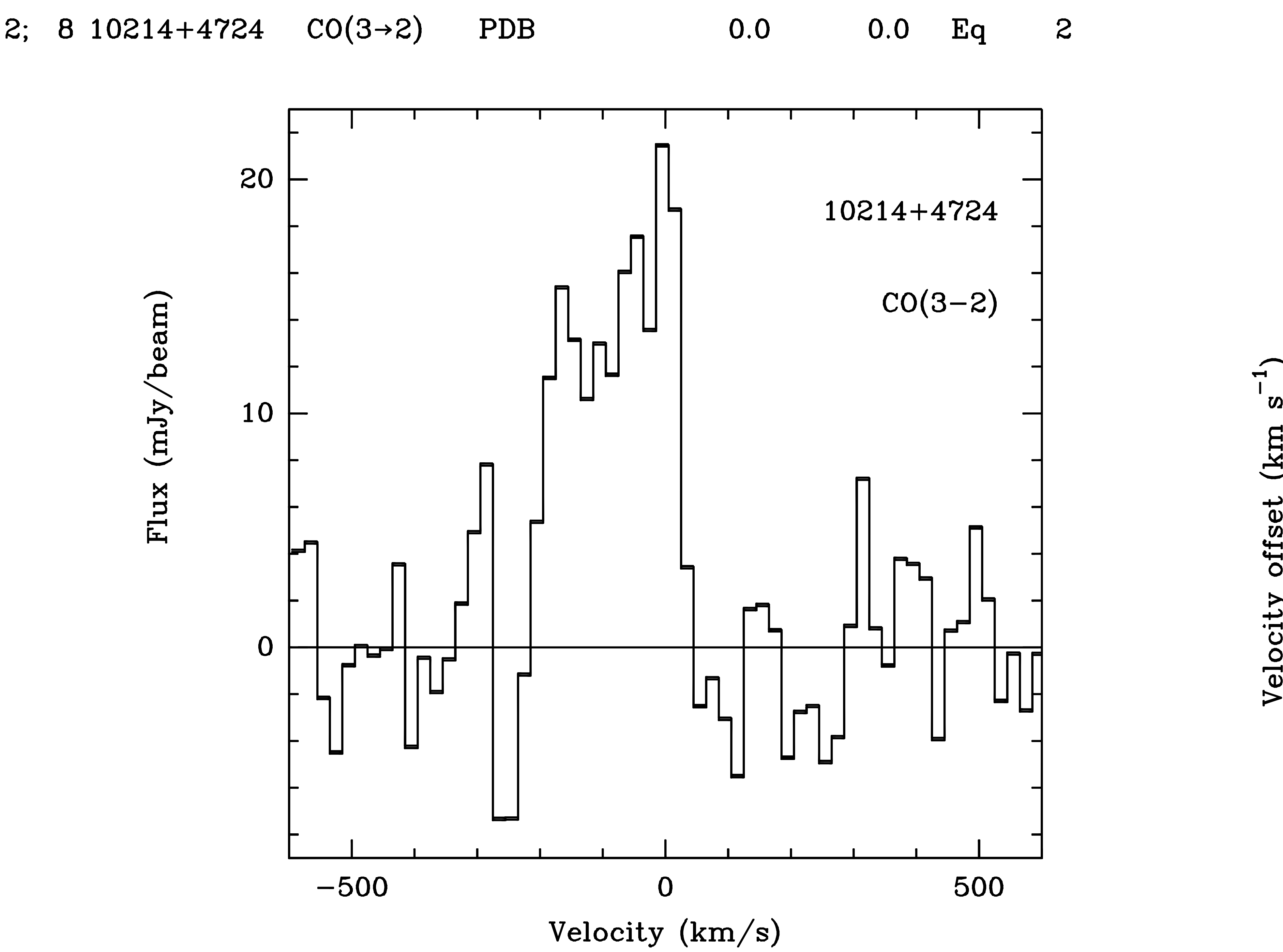
2; 8 10214+4724 CO(3→2) PDB 0.0 0.0 Eq 2
10214+4724
CO(3−2)
Flux (mJy/beam)
Velocity (km/s)
Velocity offset (km s^-1)
20
10
0
−500
0
500

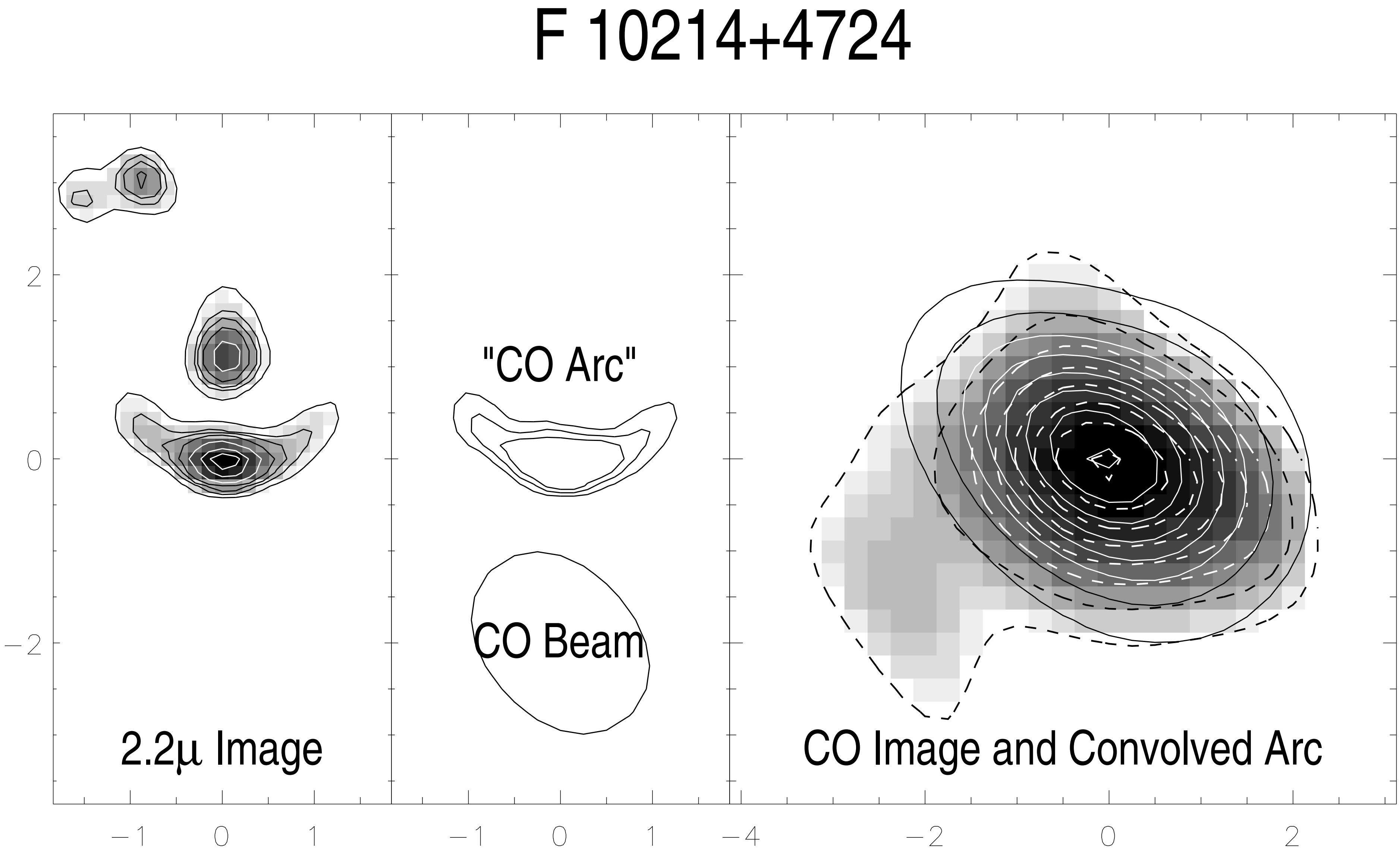